\documentclass[lettersize,journal]{IEEEtran}
\usepackage{adjustbox}
\usepackage{cite}
\usepackage{amsmath,amssymb,amsfonts}
\usepackage{amsthm}
\usepackage{algorithm}
\usepackage[caption=false,font=normalsize,labelfont=sf,textfont=sf]{subfig}
\usepackage{graphicx}
\usepackage{textcomp}
\usepackage{xcolor}
\usepackage{color,soul}
\usepackage{xspace}
\usepackage{booktabs}

\usepackage{optidef}
\usepackage{algpseudocode}
\usepackage{lipsum}
\setstcolor{red}
\usepackage{makecell}
\usepackage{graphicx}

\begin{document}
\title{Strategic Demand-Planning in Wireless Networks: Can Generative-AI Save Spectrum and Energy?}
\author{\IEEEauthorblockN{Berk Çiloğlu$^{1,2}$, Görkem Berkay Koç$^{1,2}$, Afsoon Alidadi Shamsabadi$^{2}$, Metin Ozturk$^{1,2}$, Halim Yanikomeroglu$^2$}
\IEEEauthorblockA{\\$^1$Electrical and Electronics Engineering, Ankara Yıldırım Beyazıt University, Ankara, Türkiye\\
\IEEEauthorblockA{$^2$Non-Terrestrial Networks Lab, Systems and Computer Engineering, Carleton University, Ottawa, Canada}}
\thanks{This research has been sponsored in part by The Scientific and Technological Research Council of Türkiye (TUBITAK).\\
The first three authors have contributed equally to this study.}
}

\maketitle

\begin{abstract}
Generative-AI (GenAI), a novel technology capable of producing various types of outputs, including text, images, and videos, offers significant potential for wireless communications.
This article introduces the concept of strategic demand-planning through demand-labeling, demand-shaping, and demand-rescheduling.
Accordingly, GenAI is proposed as a powerful tool to facilitate demand-shaping in wireless networks. 
More specifically, GenAI is used to compress and convert the content of various types (e.g., from a higher bandwidth mode to a lower one, such as from a video to text), which subsequently enhances performance of wireless networks in various usage scenarios, such as cell-switching, user association and load balancing, interference management, as well as disasters and unusual gatherings. 
Therefore, GenAI can serve a function in saving energy and spectrum in wireless networks. 
With recent advancements in AI, including sophisticated algorithms like
large language models and the development of more powerful hardware built exclusively for AI tasks, such as AI accelerators, the concept of demand-planning, particularly demand-shaping through GenAI, becomes increasingly relevant.
Furthermore, recent efforts to make GenAI accessible on devices, such as user terminals, make the implementation of this concept even more straightforward and feasible.
\end{abstract}
\vspace{-0.5cm}
\section{Introduction}
The growing demand for coverage and capacity, along with the shift towards the sixth-generation technology standard for mobile networks (6G), is driving the development of technologies that ensure immersive, hyper-reliable, and low-latency communication, as well as massive connectivity.
Emergence of novel use cases, aligned with the ambitious requirement of key performance indicators~(KPIs) in 6G, necessitates employment of cutting-edge technologies, such as \mbox{artificial intelligence~(AI)} and advanced network platforms, including non-terrestrial networks~(NTN)~\cite{ITU_Recommendation},\cite{NTN}.
As wireless networks become denser and various network tiers integrate vertically, forming vertical heterogeneous networks (\mbox{VHetNets}), energy consumption and resource utilization increase tremendously, requiring robust and efficient algorithms. 
Given the scale of future wireless networks and the need to optimize various metrics from both user and network perspectives, wireless networks will inevitably experience a shift towards autonomous and self-optimizing infrastructures~\cite{ITU_Recommendation}. 
This lies in the maturation of self-organization and the advanced utilization of AI technologies in future wireless networks.
The revolutionary enhancements in AI are expected to have a noticeable impact on the telecommunications field to create a sustainable, dynamic, and resilient network that can manage various conditions and scenarios efficiently~\cite{AIforWireless}. 

In recent years, AI has emerged as a promising tool for designing solutions in wireless networks.
Despite efforts to develop efficient algorithms using conventional AI and optimization methods, as well as increasing available resources (such as power, time, frequency, and hardware) through advanced technologies, including terahertz (THz) communications, and ultra-massive multiple-input multiple-output (umMIMO) antennas, proposed algorithms still fall short in scenarios where a large number of users demand a huge volume of traffic. 
Consequently, these algorithms are unable to fulfill users demand and achieve the best possible performance~\cite{DemandShaping}. 
To tackle this challenge, strategic \textit{demand-planning} can be considered a potential approach in managing the user traffic, and therefore optimizing the resource utilization and energy consumption.

\begin{figure}[t!]
    \centering
    \includegraphics[width=0.8\linewidth,trim={6cm 2cm 7cm 2cm},clip]{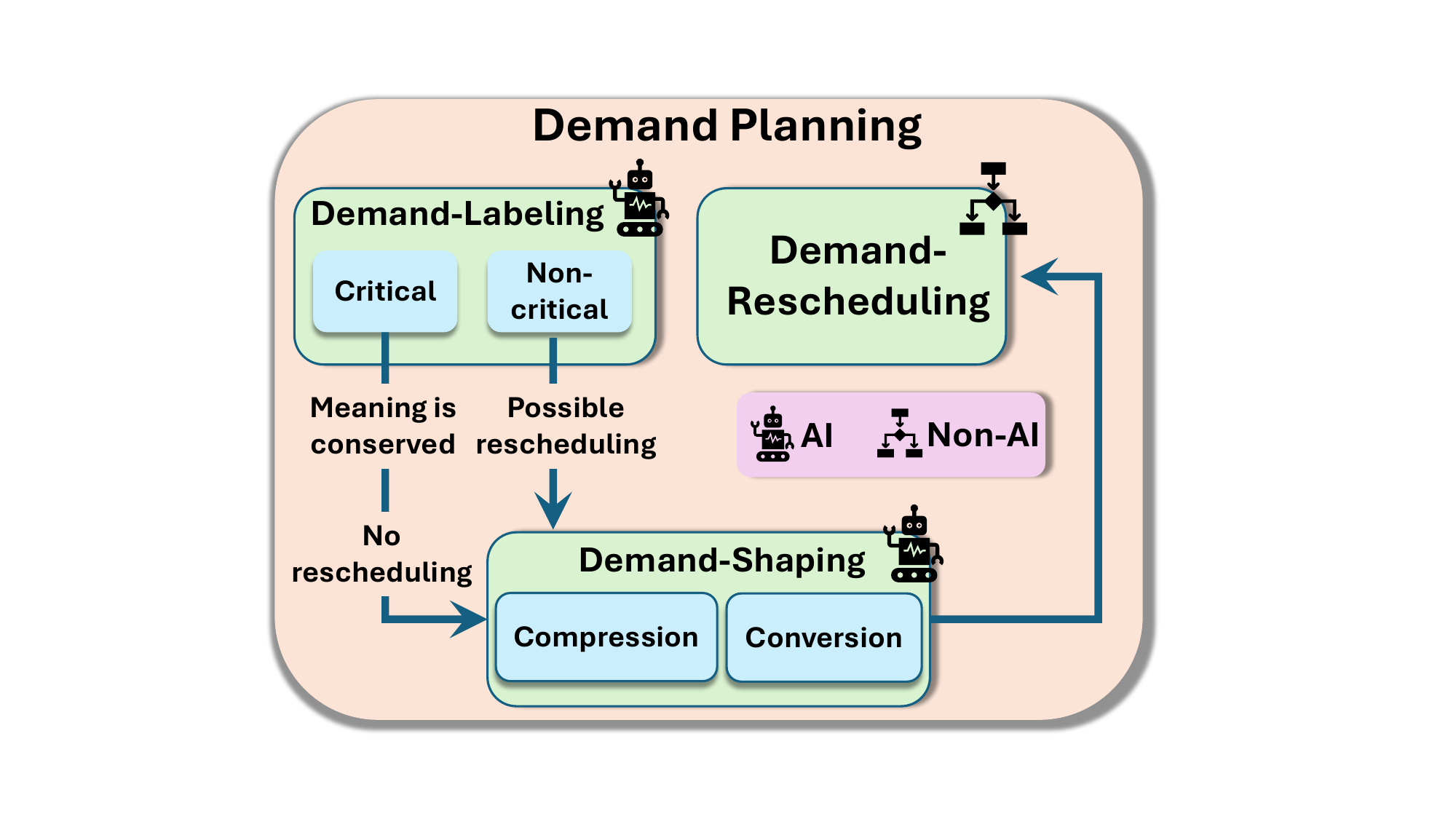}
    \caption{Demand-planning workflow in wireless networks.}
    \label{fig:demand_planning_work_flow}
\end{figure}

In wireless networks, as seen in Fig.~\ref{fig:demand_planning_work_flow}, user demand-planning involves \mbox{\textit{demand-labeling}}, \mbox{\textit{demand-shaping}}, and \textit{demand-rescheduling}. Through demand-labeling, user data can be labeled as either \textit{critical} or \textit{non-critical} based on its importance and \mbox{priority}. Accordingly, demand-shaping entails compressing user-demanded data or converting it to another format, depending on the network's resource utilization status. 
In scenarios of network congestion, where available resources are insufficient to fulfill user's requirement, the user data that needs to be sent with high priority and can be shaped without meaning change or information loss, can be shaped, and transmitted using fewer allocated resources. 
On the other hand, user data with a lower priority level or requiring precise transmission of bits can be rescheduled to a time slot where available resources are sufficient for reliable transmission. 

Recently, there has been a noticeable interest in employing \textit{generative-AI (GenAI)} models in the telecommunications field. To this end, several studies have explored the potential of utilizing GenAI for network planning and optimization~\cite{GenAIfor5G},\cite{DGN}. Furthermore, GenAI has been considered a promising tool for unveiling new technological possibilities for 6G~\cite{genai_telecom},\cite{GenAI_PhyLayer}. 
Besides the conventional applications of GenAI in traditional communications that guarantee the accurate transmission and reception of the user's data, GenAI has the potential to play a crucial role in semantic communications\cite{SemCom-ProfHalim}. Additionally, GenAI can be utilized at the user terminal or base stations (BSs) to facilitate demand-planning in wireless networks in scenarios where the available resources are insufficient to accommodate the entire demanded traffic. 
Through demand-shaping, GenAI improves the performance of wireless networks from resource utilization and energy consumption perspectives, i.e., energy and spectrum are saved. 
Therefore, demand-planning, and particularly demand-shaping through GenAI, can be a high-potential \mbox{remedy} to enhance future ultra-dense wireless networks, making them more sustainable, dynamic, and resilient.
\section{Demand-Planning with GenAI: Concept and Models}\label{Gen-AI Concept}
GenAI technology has undergone rapid enhancements, leading to the development of novel generative models. Each GenAI model comprises various components, and has been designed to achieve a specific objective. In this section, we first discuss the basics of GenAI along with the key models. Next, we introduce the concept of demand-planning, explore its application in wireless networks, and discuss the associated advantages.
\vspace{-0.3cm}
\subsection{GenAI Basics}
GenAI is an AI technology capable of generating original and intricate content, including texts, images, videos, etc., using patterns and information, learned from extensive datasets\cite{genai_survey}. 
GenAI plays an important role in various fields, such as data augmentation, sentiment analysis, question-answering, conversational interfaces, and automation, by improving human-machine interactions and generating new content. 
Additionally, GenAI is a potential tool for optimizing the performance of future wireless networks, such that it
can improve data optimization with the help of highly efficient and versatile models. 
Among these, generative adversarial networks (GANs) are used for adversarial training; variational autoencoders (VAEs) for probabilistic encoding; diffusion models (DMs) for data diversity; and transformers for sequence processing~\cite{genai_telecom}.

Two competing neural network models exist in GANs, engaged in a zero-sum game, i.e., if one model gains, the other loses.    
The first neural network, referred to as \textit{generator}, generates fake data samples (i.e., synthetic data) based on the training set, and the second one, named \textit{discriminator}, determines how realistic the fake data seems. 
These two parts work together to generate new data with the same statistics as the training set. GAN models have been widely used in the wireless networks literature to analyze transmitted data, and to model traffic patterns and user behaviors. This supports wireless networks in estimating network demand and required resources more efficiently. VAEs are another type of GenAI approaches, that are generative models created by combining deep learning and probabilistic graphic models. VAEs are used to learn complex probabilistic distributions from data sets and generate new data samples. 
Therefore, they are potential algorithms for anomaly detection and data compression. 
In particular, VAEs can be instrumental in user's demand-planning in future wireless networks, enabling the optimization of the resource utilization and energy consumption, where the available resources are not sufficient.
\begin{table}[]
\caption{GenAI models comparison}\label{table:GenAI_models}
\vspace{-0.1cm}
\resizebox{\columnwidth}{!}{
\begin{tabular}{lll}
\hline
\textbf{Model Type}   & \textbf{Fundamental Concepts}  & \textbf{Applications}             \\ \hline
\textbf{GANs}         & \begin{tabular}[c]{@{}l@{}}Consists of two neural\\ networks: \textit{generator} creates\\
data,
and the \textit{discriminator}\\
evaluates it.\end{tabular}   & \begin{tabular}[c]{@{}l@{}}- Channel modeling \\ \hspace{0.2cm}and simulation. \\ - Generating synthetic \\\hspace{0.2cm}network traffic for testing. \end{tabular} \\ \hline
\textbf{VAEs}         & \begin{tabular}[c]{@{}l@{}}Utilizes a probabilistic \\approach to understand \\data distribution.\end{tabular}           & \begin{tabular}[c]{@{}l@{}}- Anomaly detection in \\\hspace{0.2cm}network traffic. \\- Network performance\\\hspace{0.2cm}optimization.\end{tabular}                                           \\ \hline
\textbf{DMs}          & \begin{tabular}[c]{@{}l@{}}Gradually adds and then \\removes noise from data \\to generate new samples.\end{tabular} & \begin{tabular}[c]{@{}l@{}}- Noise reduction in \\\hspace{0.2cm}communication signals. \\- Enhancing data clarity \\\hspace{0.2cm}in low-quality signals.\end{tabular}                          \\ \hline
\textbf{Transformers} & \begin{tabular}[c]{@{}l@{}}Processes data in parallel \\and captures dependencies \\over sequences.\end{tabular}           & \begin{tabular}[c]{@{}l@{}}- Enhancing customer \\\hspace{0.2cm}service. \\- Adaptive optimization \\\hspace{0.2cm}of network data.\end{tabular}            
\\ \hline
\vspace{-0.8cm} 
\end{tabular}
}
\end{table}
\begin{figure*}[ht!]
    \centering
    \captionsetup{justification=centering}
    \includegraphics[width=0.72\linewidth,trim={0cm 0cm 0cm 0cm},clip]{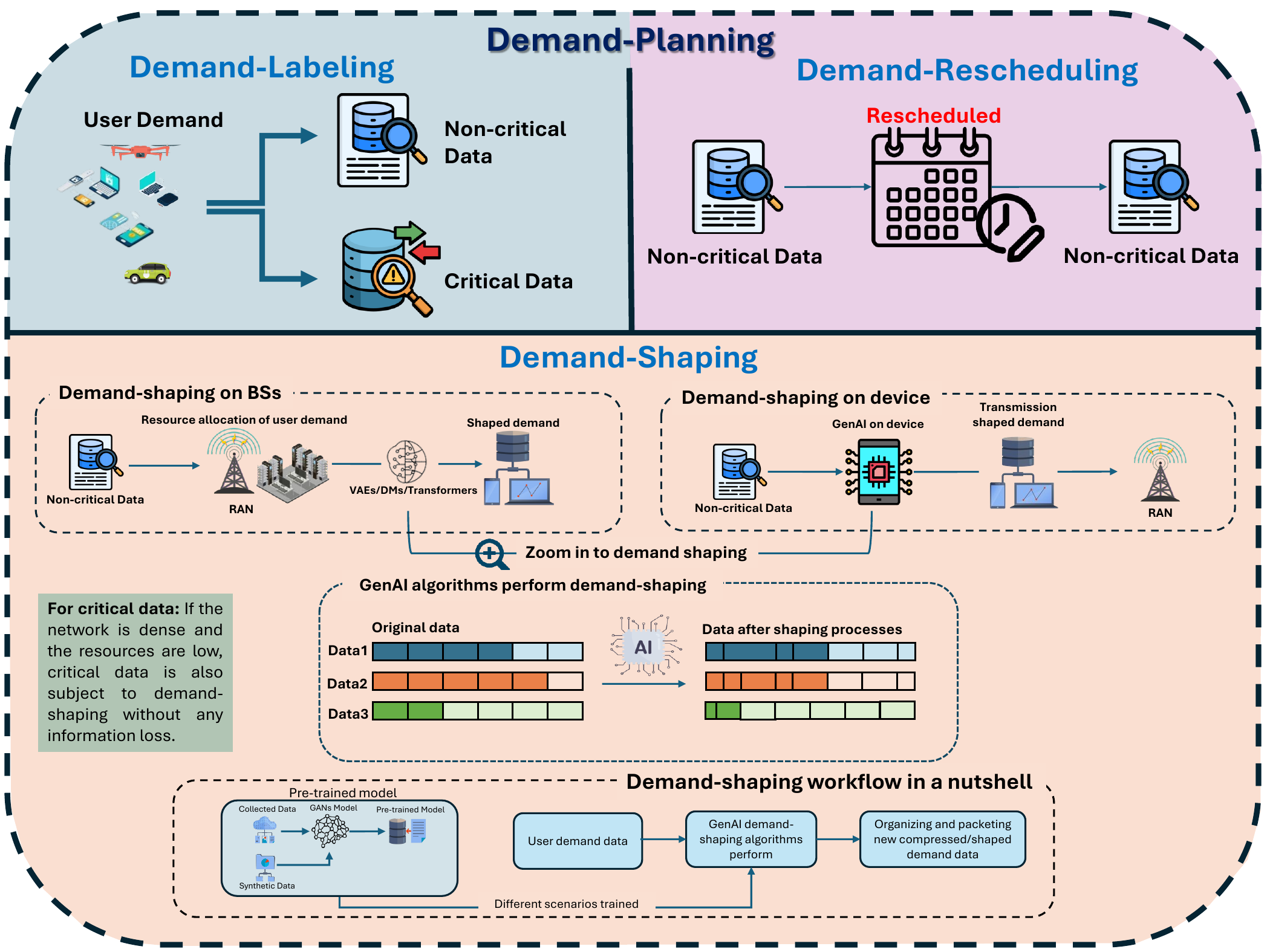}
    \caption{GenAI demand-planning concept model and methodology.}
    \label{fig:Gen-AI_demand_shaping}
    \vspace{-0.3cm}
\end{figure*}

DMs are two-stage deep learning models: in the first stage, noise is incrementally added to real data samples until the data is transformed into complete noise, while in the second stage, the model reduces this noise step-by-step, approaching the original data. 
In this way, the model learns how to generate realistic data from random data by removing the noise. GenAI DMs are essential in noise reduction and traffic prediction in wireless networks.
Transformers, on the other hand, capture long-term dependencies and contextual information within sequences, making them particularly effective for data compression and data optimization tasks.
Furthermore, compared to conventional sequential models, transformers can process entire sequences in parallel, leading to faster and more efficient optimization processes. 
It is worth mentioning that large language models (LLMs) are trained and developed using transformer architectures and have potential applications in telecommunications networks~\cite{genai_telecom,SemCom-ProfHalim}. 
LLMs can optimize data and network performance by prioritizing critical information and adapting to contextual changes. Additionally, they can predict data sequences, enabling the reconstruction of lost data, which ultimately enhances the reliability of communications.
The comparisons of GenAI models are summarized in Table~\ref{table:GenAI_models}.

It is worth noting that major manufacturers, such as Apple, have started forming partnerships with AI companies for on-device GenAI usage~\cite{openai-apple-partnership}, thereby proving the applicability of the concept introduced in this work.
\vspace{-0.2cm}
\subsection{Demand-Planning through GenAI}
The primary objective of demand-planning in wireless networks is to reorganize user requests to minimize resource utilization (e.g., energy and spectrum) as much as possible. 
To facilitate this, a plug-in software incorporating GenAI models can be installed on user terminals or at BSs to perform demand-planning in uplink or downlink communications. 
As illustrated in Fig. \ref{fig:demand_planning_work_flow}, demand-planning in wireless networks is implemented through demand-labeling, demand-shaping, and demand-rescheduling processes.
\subsubsection{Demand-labeling}
Considering the importance and priority of user data, it can be categorized as either critical or \mbox{non-critical}. 
Critical data requires high-priority transmission, while non-critical data can be sent with lower priority. 
In cases of insufficient network resources, critical data transmission may encounter significant challenges. In such scenarios, \mbox{non-critical} data can be rescheduled for transmission at a later time, while critical data, whose meaning is supposed to remain unaffected, can be processed by demand-shaping algorithms and transmitted with fewer required resources. It is noteworthy that the critical data, which cannot be shaped due to the information loss risk, should be transmitted promptly using the available resources.
\begin{figure*}[t!]
    \centering
    \captionsetup{justification=centering}
    \includegraphics[width=.8\linewidth,trim={1cm 0.6cm 0.7cm 0cm},clip]{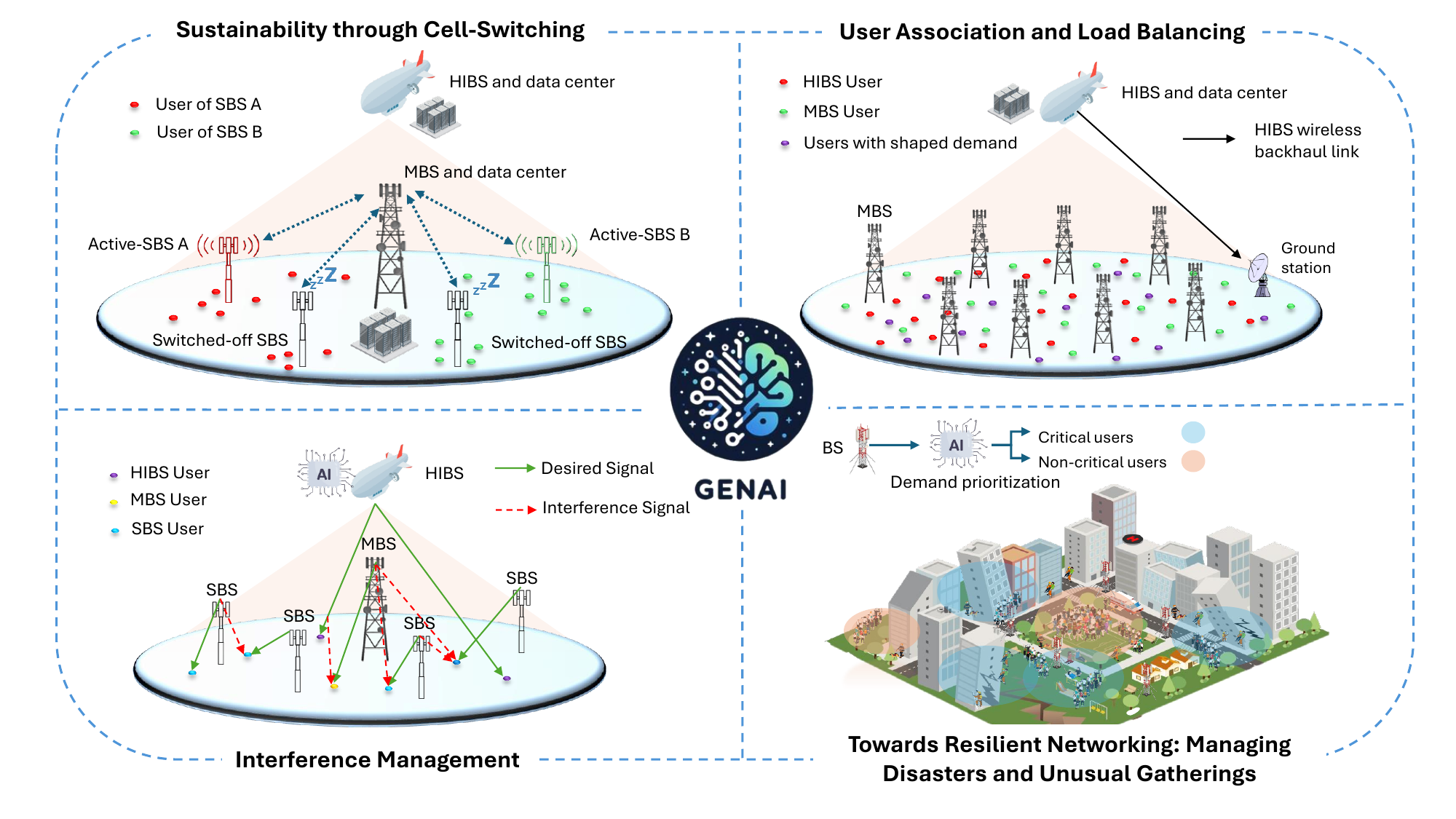}
    \caption{GenAI demand-shaping usage scenarios in wireless networks. The GenAI logo in the middle of the figure has been created by the DALL·E 2.}
    \label{fig:Gen-AI_Applications}
    \vspace{-0.2cm}
\end{figure*}
\subsubsection{Demand-shaping}
With the help of the aforementioned models, GenAI can compress or convert the user data to optimize the network's traffic through demand-shaping. 
Data compression can support shaping user data into a lower volume, which requires lower data rates and fewer allocated resources to be transmitted.
For instance, the size of high-resolution videos can be reduced through video compression algorithms provided by GenAI at the transmitter, and the shaped video can be converted back to its original format (i.e., decompression) using DMs at the receiver side. To achieve the optimal performance, the GenAI models at the transmitter and receiver should be consistent.
While any compression algorithm can reduce data size, GenAI excels with its content-aware and regenerative capabilities. It analyzes data to identify key sections and applies variable compression based on their importance, compressing less critical parts more heavily. Additionally, it can regenerate content to correct any compression-related errors. 

Data conversion is another approach to perform demand-shaping in wireless networks. In particular, contents such as multimedia and video, which require relatively more resources, can be converted to a lower-data-rate format before being transmitted through the network. Another way to shape the demand can be by analyzing and optimizing the content of the data.
In other words, GenAI can make comprehensions from the meaning of the data, and optimize the content.
To this end, voice and text messages can be summarized using GenAI, without loss of inference from its original type, to compress the user data to further relax the network traffic.
\subsubsection{Demand-rescheduling}
Although demand-shaping is a powerful tool for managing network traffic, applying this process is not feasible for certain types of data due to the risk of information loss. To address this issue, demand can be rescheduled, meaning that content can be transmitted at a different time, such as when data traffic is idle or relatively low. Furthermore, if the network's available resources are insufficient to accommodate both critical and non-critical data, demand-rescheduling can be a potential solution for non-critical data to decrease the network traffic load.

Fig. \ref{fig:Gen-AI_demand_shaping} illustrates the concept and methodology of demand-planning through GenAI in wireless networks. 
The process begins with user data labeling and proceeds to demand-shaping. 
Pre-trained GAN models generate synthetic data to complement the collected data in demand-shaping; therefore, a comprehensive dataset is ensured. 
The user demand can then be compressed and optimized using VAEs, DMs, or transformer algorithms. 
Hence, the resource allocation and network performance become more efficient thanks to the optimized dataset as well as the compression process.
Additionally, the transmission of non-critical data can be rescheduled when necessary.
In conclusion, due to the gap in a single GenAI model being able to perform these tasks in the literature, we envision using different GenAI models to perform various tasks. When these models are used together, they can execute the desired demand planning operations.
\vspace{-0.25cm}
\section{Usage Scenarios in Wireless Networks}\label{applications}
Employing GenAI models for demand-planning in wireless networks provides advantages in various usage scenarios. 
Particularly, demand-planning can be implemented in uplink or downlink communications to improve the resource utilization of the wireless networks, in scenarios where the resources are insufficient to fulfill the users' demand. 
According to the network's available resources and specific requirements of the users, a GenAI model can be employed, at the user terminal and/or the network side to facilitate demand-planning.

In this section, we introduce potential usage scenarios, where demand-planning through GenAI can improve the performance of wireless networks in terms of energy consumption and spectral efficiency. 
These usage scenarios, illustrated in Fig. \ref{fig:Gen-AI_Applications}, include sustainability through cell-switching, user association and load balancing, interference management, and disasters and unusual gatherings that necessitate quick decision-making.
Each usage scenario contains a diverse set of users and BS types, such as small BS~(SBS), macro BS~(MBS), and high altitude platform stations (HAPS) as international mobile telecommunications (IMT) BS (HIBS).
It is worth mentioning that the GenAI logo at the center of Fig. \ref{fig:Gen-AI_Applications} has been created by the DALL·E, a text-to-image model developed by OpenAI. 
\vspace{-0.44cm}
\subsection{Sustainability through Cell-Switching}\label{sec:usage_cell_switch}
As wireless networks' energy consumption increases, so does the amount of carbon released into the atmosphere, which is causing a failure to meet climate change commitments, one of the most significant challenges of our time. One important technology for reducing energy consumption in wireless networks is cell switching, a viable solution that seeks idle or lightly loaded SBSs, offloads their traffic to MBSs, and puts them into sleep mode as they are no longer needed to operate~\cite{haps_cs_ours_trans}.
The operating principle of cell-switching is based on the density of the resource blocks (RBs) of BSs (i.e., data traffic loads).
GenAI-based demand-planning allows for adjusting the RBs demanded by each user, thereby optimizing the resource utilization of a BS.
Hence, the available capacity of the MBS
increases due to the lower amount of data offloaded from switched-off BSs, as a result of demand-planning process, providing more switching off opportunities.
Consequently, the energy consumption of the overall wireless network will be decreased, leading to a more sustainable networking. 

\begin{figure}[t!]
    \centering
    \includegraphics[width=0.87\linewidth]{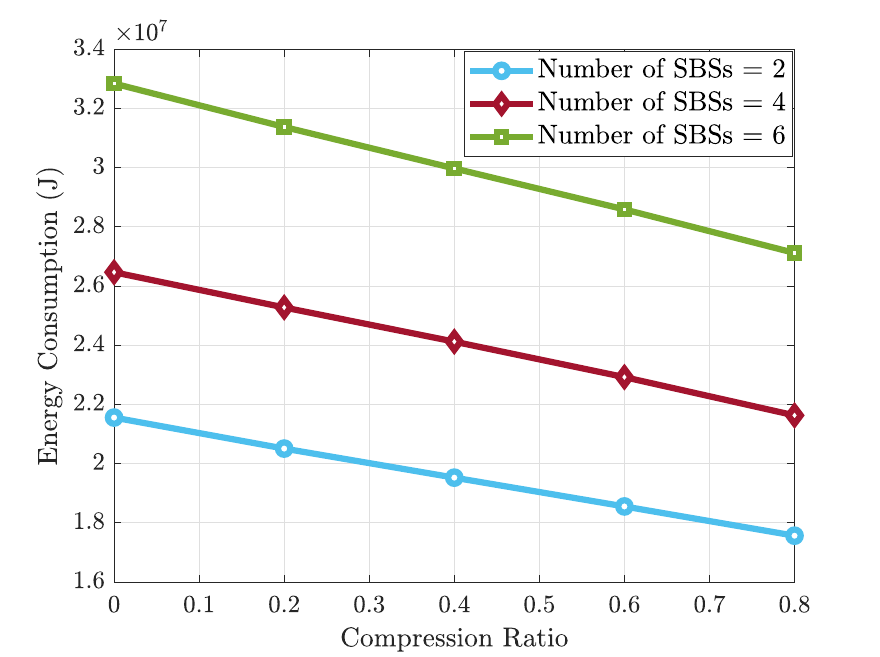}
    \caption{Impact of demand-shaping through GenAI on energy consumption in implementing cell-switching in a wireless network.}
    \label{fig:CS_compress_demand}
    \vspace{-0.2cm}
\end{figure}

The impact of demand-shaping through GenAI on the energy consumption of wireless networks is depicted in Fig. \ref{fig:CS_compress_demand}. 
For this purpose, the daily data demands of the users from the city of Milan, collected by Telecom Italia~\cite{milan}, are considered and cell-switching is performed with an exhaustive search algorithm in networks with diverse set of BSs (1 MBS, and 2, 4, 6 SBSs). 
We consider that the actual user demand is compressed by GenAI using various compression ratios: 0\%, 20\%, 40\%, 60\%, and 80\%.
It is observed from the findings in Fig.~\ref{fig:CS_compress_demand} that by increasing the compression ratio of demand-shaping algorithm, the network's total energy consumption decreases.
In more complex and dense networks, consisting of NTN platforms (e.g., HIBS~\cite{haps_cs_ours_trans}), demand-shaping will play a substantial role in optimizing energy consumption. For instance, after shaping the demands through GenAI, NTN elements like HIBS can host the offloaded users, increasing more switching-off possibilities that is translated into even more energy saving as HIBS is envisioned to be empowered with green-friendly energy sources.
An important point to consider is the trade-off between the energy savings achieved through GenAI-assisted cell switching and the increased energy consumption by GenAI algorithms. 
The increased energy consumption of AI/GenAI implementations presents a challenge in terms of carbon emissions that must be addressed during the design process to ensure that the energy savings significantly outweigh the energy consumed by the AI.
\vspace{-0.33cm}
\subsection{User Association and Load Balancing}\label{sec:usage_load}
Designing efficient user association schemes is a crucial challenge in wireless networks, particularly in VHetNets where various network tiers, including MBSs, SBSs, and aerial platforms, collaborate to meet users demands. User association schemes play a vital role in designing dynamic and resilient algorithms to enhance the network performance and balance the load among various tiers. However, the constraints posed by the network available resources as well as the immersive traffic demand of the users restrict the flexibility in designing the most efficient schemes. 

Among these limitations is the restricted wireless backhaul link capacity of aerial platforms, such as HIBS, or the terrestrial BSs lacking fiber facilities. Due to the limited available spectrum resources and the backhaul link capacity constraint, BSs are limited in the number of users they can support. By implementing GenAI at BSs with wireless backhaul capacity, user data can be compressed to a format that requires less amount of allocated spectrum, and a lower data rate, preventing congestion on the wireless backhaul link. To this end, the BS analyzes the users' demanded traffic, followed by deciding to implement the GenAI algorithm to compress the users' data based on available capacity and spectrum in the wireless backhaul link. Considering the type of users' data, the BS may also decide to select a specific group of users for demand-shaping.

Another challenge in wireless networks involves the uplink communication from users to BSs. Given the limited transmit power and battery energy storage of the user equipment, managing uplink communication in scenarios characterized by significant free-space path loss is a critical challenge. In these situations, considering the data type and the available energy resources at the user side, GenAI can be implemented at the user terminal to compress the demanded data, enabling it to be transmitted using less transmit power.
\vspace{-0.3cm}
\subsection{Interference Management}\label{sec:usage_interference}
\begin{figure}[t!]
    \centering
    \includegraphics[width=0.86\linewidth]{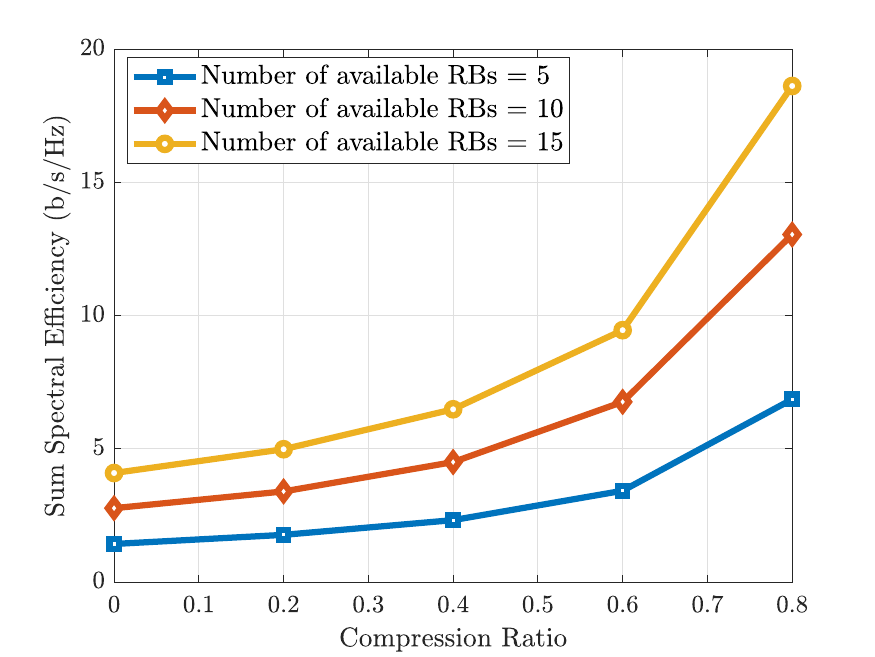}
    \caption{Impact of demand-shaping through GenAI on sum spectral efficiency in a VHetNet consisting of 1 HIBS and 4 MBSs, serving 50 users.}
    \label{fig:IM_compress_demand}
    \vspace{-0.2cm}
\end{figure}
VHetNets enhance the coverage and capacity of terrestrial networks and contribute to making wireless networks more resilient. 
Nevertheless, ultra-dense VHetNets, particularly those operating in a harmonized spectrum, where various network tiers share the same frequency band, experience performance issues due to propagated inter/intra-tier interference~\cite{IM-RA2}. 
Consequently, designing effective interference management mechanisms becomes crucial. 

One of the most common approaches to optimize interference in wireless networks is through designing robust and efficient spectrum allocation schemes\cite{IM-RA2}.
However, despite the efforts towards developing efficient spectrum allocation algorithms, the continuous rise in traffic demands from mobile users will be an obstacle. This means that in certain scenarios, the interference management algorithms, developed through spectrum allocation schemes, fail to deliver the expected improvements, specifically due to the limited amount of available spectrum resources. Demand-planning through GenAI increases the degree of freedom in allocating spectrum to users. Employing GenAI demand-planning, BSs can shape the user data and manage traffic more effectively, leading to enhanced interference management. Fig.~\ref{fig:IM_compress_demand} represents the sum spectral efficiency in a VHetNet, consisting of 1 HIBS and 4 MBSs, where the total available RBs are shared by 50 users. As depicted in Fig.~\ref{fig:IM_compress_demand}, by compressing the demanded data by the users, the RBs can be allocated to the users more efficiently, leading to decreased interference and increased sum spectral efficiency.
\vspace{-0.3cm}
\subsection{Towards Resilient Networking: Managing Disasters and Unusual Gatherings}\label{sec:usage_resillient}
In disaster scenarios, it is crucial for search and rescue~(SAR) teams to reach users quickly and make expedited decisions. Therefore, the communication channel between the SAR teams and users, as well as among the SAR teams themselves, must be reliable and fast to ensure effective support for affected individuals. Nevertheless, in disaster scenarios, telecommunication infrastructures are also affected and the availability of reliable links gets restricted, leading to a decreased amount of available resources. In such scenarios, the demand-planning through
GenAI can be a potential solution to enhance the efficiency of communication channels, enabling more number of users to be served. 

To this end, user-generated content such as multimedia and text can be initially categorized as either ``critical" or ``non-critical" in the demand-labeling step, according to their importance to the SAR mission.
Critical content requires faster and more reliable communication, and should be delivered with high priority.
On the other hand, if the communication traffic between two nodes in the affected area is not related to the disaster situation, it can be categorized as ``non-critical". During the disaster scenarios, in case of lack of adequate available resources to transmit the critical data, the non-critical data can be rescheduled to another time (when data traffic is relatively low, and the resources are sufficient to accommodate the traffic), and the critical data, that their meaning will not be affected by demand-shaping, can be shaped by GenAI models to require less resources for transmission. 
It is worth highlighting that although demand-shaping may add additional computation and processing delay, the compressed data will be transmitted with less propagation delay.
To this end, demand-shaping through GenAI will reduce data traffic, alleviating the load on communication networks, and addressing capacity issues in a disaster scenario. 

Apart from its role in disaster scenarios, GenAI can build forecasting models for network traffic demand, accounting for user behavior-influencing factors, such as the time of day and special occasions.
Prior to a major public event, for example, GenAI can simulate the expected increase in demand to ensure that, allowing the network designers to create and analyze models resulting from this simulation to proactively plan necessary infrastructure upgrades or adjustments before the demand peaks occur.
Moreover, during large public events, GenAI can manage network density by distributing demand to BSs or compressing/converting user-generated content.
\vspace{-0.3cm}
\subsection{Use-Cases and Business Implications}
The usage scenarios, enabled by the proposed GenAI-based demand shaping framework, have the potential to facilitate new use-cases with significant business implications. The aforementioned usage scenarios
support network operators in managing the network with the existing available spectrum and energy resources. Considering that the spectrum is an expensive asset for the operators, this significantly supports them in saving the cost of acquiring additional spectrum.
Additionally, the remnant spectrum that is saved with the proposed approaches can be repurposed, such as leasing it to other operators. 
In this way, primary operators can increase their profits through a new revenue stream from spectrum leasing, while secondary operators gain more opportunities to enhance the quality of service~(QoS) for their users.

Furthermore, the energy savings from BS switching are reflected in reduced operational costs, making the businesses of operators more economically sustainable.
The more BSs that can be switched off, the greater the potential impact. GenAI-based demand shaping plays a crucial role here, as it enables the switching off of more BSs by reducing their load. In addition, the usage scenario described in Section~\ref{sec:usage_resillient} is a critical use case where the proposed methodology can save lives during disaster situations. Moreover, it plays a vital role in unusual gatherings by improving or maintaining user satisfaction, a key factor in business sustainability.
\vspace{-0.2cm}
\section{Challenges and Open Issues}\label{challenges}
According to the discussions in Sections \ref{Gen-AI Concept} and \ref{applications}, it is evident that GenAI will be an indispensable part of the future of wireless networks. 
On the flip side, implementing GenAI in wireless networks
has its own challenges, and in the following paragraphs, we highlight some of the key challenges.
\vspace{-0.3cm}
\subsection{Hardware Bottlenecks and Data Content Access}
Although BSs are equipped with a relatively more powerful hardware architecture, arming user equipment with GenAI might be a bottleneck in realizing GenAI-based demand-shaping at the device level. Furthermore, the BS faces challenges in applying GenAI due to its lack of access to application-layer data. Edge computing offers a solution by handling content analysis and AI tasks at nearby nodes, reducing the processing burden on the BS.

\vspace{-0.3cm}
\subsection{Regulatory Policies and Carbon Emission}
Regulatory policies, such as net neutrality, aim to maintain a fair data service for all the data, regardless of its source, content, or destination. Demand-planning through GenAI can theoretically be designed to support net neutrality by ensuring that all users with the same service-level agreements (SLAs) receive equal treatment. At the same time, the GenAI should not prioritize traffic unfairly or violate these agreements. In any scenario, prioritizing demand planning during emergencies can be justified ethically and legally to protect public safety, provided such actions are temporary.
Demand-shaping through GenAI requires high computation and processing capabilities to compress the data, boosting the energy consumption, and therefore the carbon emission which is a critical issue.
GenAI algorithms are applied in demand shaping for good reasons, but the increased carbon emissions resulting from these computationally demanding algorithms should not overshadow the benefits achieved in demand shaping.
\vspace{-0.3cm}
\subsection{Cyber Security and Encrypted Data}
During the demand-shaping process, there is a risk of malicious interventions aimed at disrupting the system. In addition, data can be accessed, monitored, or manipulated by malicious parties, raising serious privacy concerns. Such abuses not only threaten system functionality but also highlight the need for robust security measures.
The research highlighted in~\cite{stanford_ai_index} underscores the importance of addressing security vulnerabilities, emphasizing that AI can potentially increase cyber attacks.
Another difficulty for GenAI in performing demand-planning operations is its approach to encrypted data. 
Therefore, while GenAI is promising in processing various types of data, encrypted data traffic requires further research to ensure both efficiency and security in future wireless networks.
\vspace{-0.5cm}
\subsection{Information Loss and QoS Degradation}
A common flaw in
data shaping is the oversimplification of information. Although GenAI algorithms generally perform with good accuracy, there is still a risk of losing contextual information during the demand-shaping process. This can result in the omission of important details and nontrivial nuances, leading to incomplete information for the receiver. Consequently, the SLA and QoS of the users will be affected. Therefore, it is crucial to have efficient mechanisms to prevent information loss and maintain the QoS of the users.
In scenarios where information loss is unavoidable, data rescheduling is a potential replacement solution for demand-shaping.
\vspace{-0.3cm}
\subsection{Consistency, Data Privacy, and Transparency}
GenAI models can sometimes produce inconsistent outputs for the same content, leading to unreliable or unpredictable results. This is especially critical in situations such as disasters, where models need to operate smoothly and quickly. Therefore, it is essential to develop and continuously improve validation methods to enhance the reliability of GenAI models.
Besides, when processing user demands, the security and confidentiality of data are of high importance. 
Additionally, the \textit{black box} nature of GenAI, with its lack of transparency about how algorithms work, can make it difficult for users to trust the process.
\vspace{-0.4cm}
\section{Conclusion}\label{Conclusion}
This paper presents the novel demand-planning approach through GenAI technology and discusses its potential usage scenarios in wireless networks, particularly 6G which is expected to be AI-oriented.
The underlying idea in this article is to manage the network data traffic through demand-labeling, demand-shaping, and demand-rescheduling. 
To this end, GenAI is proposed as the tool to implement demand-shaping and facilitate various applications, including cell-switching, user association and load balancing, interference management, and disaster scenarios, with the aim of saving energy and spectrum.
The demand-shaping approach can be implemented at the user terminals and/or BSs, considering the available resources in uplink and downlink communications. 
The usage scenarios are primarily discussed within the scope of the integration of non-terrestrial and terrestrial networks, as NTN is envisioned to be the new frontier in wireless communication networks, especially for ubiquitous connectivity.
\vspace{-0.3cm}
\bibliographystyle{IEEEtran}
\bibliography{output}
\section*{Biography}
\vspace{-35pt}
\begin{IEEEbiographynophoto}{Berk Çiloğlu}
    (cilogluberk@gmail.com) is a M.Sc. student of the EMIMEP at Université de Limoges, Limoges, France. 
    His research focuses on applications of 6G and beyond wireless communication technologies, and non-terrestrial networks.
\end{IEEEbiographynophoto}
\vspace{-35pt}
\begin{IEEEbiographynophoto}{Görkem Berkay Koç}
    (gorkembrkoc@gmail.com) is a M.Sc. student at Ankara Yıldırım Beyazıt University, Ankara, Türkiye.
    His research interests focus on optimizing wireless network performance, specifically non-terrestrial networks, and visible light communications.
\end{IEEEbiographynophoto}
\vspace{-35pt}
\begin{IEEEbiographynophoto}{Afsoon Alidadi Shamsabadi}
(afsoonalidadishamsa@sce.carleton.ca) is a Ph.D. Candidate in the CU-NTN Lab at Carleton University, Canada. Her research focuses on the innovative integration of NTN with terrestrial wireless networks.
\end{IEEEbiographynophoto}
\vspace{-35pt}
\begin{IEEEbiographynophoto}{Metin Ozturk} (metin.ozturk@aybu.edu.tr) is 
Assistant Professor at Ankara Yıldırım Beyazıt University, Turkiye.
His current research includes intelligent networking for wireless communications, with a particular focus on NTN and sustainability.
\end{IEEEbiographynophoto}
\vspace{-35pt}
\begin{IEEEbiographynophoto}{Halim Yanikomeroglu} (halim@sce.carleton.ca) is a Chancellor's Professor in the Department of Systems and Computer Engineering at Carleton University, Ottawa, Canada, leading the CU-NTN Lab. His research group focuses on wireless access architecture for the 2030s and 2040s, with a particular emphasis on NTN.
\end{IEEEbiographynophoto}
\end{document}